\documentclass[review,number,sort&compress]{elsarticle}
\usepackage{amssymb}
\usepackage{amsmath}
\usepackage{color}
\usepackage[normalem]{ulem}
\usepackage{array, booktabs, lscape, rotating}
\usepackage[a4paper]{geometry}

\journal{Nuclear Instruments and Methods in Physics Research A}

\begin{document}

\begin{frontmatter}

\title{Preliminary study of feasibility of an experiment looking for excited 
state double beta transitions in Tin}

\author{Soumik Das$^{1,2}$, S. K. Ghorui$^{2,}$\footnote{Corresponding
author. {\it E-mail address:} surja@iitrpr.ac.in
(S.~K.~Ghorui).}, P. K. Raina$^2$,  
A. K. Singh$^1$, P. K. Rath$^3$,  F. Cappella$^4$, R. Cerulli$^4$,  
M. Laubenstein$^4$, P. Belli$^5$, R. Bernabei$^{5,6}$}

\address{
$^1$Dept. of Physics, Indian Institute of Technology Kharagpur, Kharagpur, IN-721302, India\\
$^2$Dept. of Physics, Indian Institute of Technology Ropar, Ropar, IN-140001, India\\
$^3$Dept. of Physics, University of Lucknow, Lucknow, IN-220628, India \\
$^4$INFN, Laboratory Nazionali del Gran Sasso, I-67100 Assergi (AQ), Italy\\
$^5$INFN sezione Roma ``Tor Vergata", I-00133 Rome, Italy \\
$^6$Dipartimento di Fisica, Universit$\grave{a}$ di Roma  ``Tor Vergata",  I-00133 Rome, Italy\\}

\begin{abstract} 
An attempt to study the feasibility of a new experiment to search for double beta decay in
$^{112}$Sn and $^{124}$Sn was carried out by using ultra-low background HPGe detector (244 cm$^{3}$)
inside the Gran Sasso National Laboratory (LNGS) of the INFN (Italy). A small sample of 
natural Sn was examined for 2367.5 h. The radioactive contamination of the sample has 
been estimated. The data has also been considered to calculate the present sensitivity for 
the proposed search; half-life limits $\sim$ $10^{17} - 10^{18}$  years  for $\beta^{+}$EC 
and EC-EC processes in $^{112}$Sn  and $\sim$ $10^{18}$ years for $\beta^{-}\beta^{-}$  
transition in $^{124}$Sn were measured. In the last section of the paper the enhancement of 
the sensitivity for a proposed experiment with larger mass to reach theoretically estimated 
values of half-lives is discussed.

\end{abstract}


\end{frontmatter}

%\linenumbers

\section{Introduction}

 The existence of non-zero mass of neutrino has been established
by neutrino flavor-oscillation experiments \cite{kamio, sud, kamla}. 
However, the nature of the neutrino, either
Dirac or Majorana, can only be tested through observation of neutrino-less
double beta decay, which is a rare second order transition involving two isobars.
Moreover, the double beta decay  experiments have the potential to establish the absolute
scale of the neutrino mass, to prove the hierarchy of neutrino mass, to test
the existence of right-handed admixtures in the weak interaction, and to test
some other effects beyond the Standard Model. The 0(2)$\nu\beta\beta$ decay can occur as :
\begin{eqnarray}
(A,Z)\rightarrow(A,Z+2) +2e^{-} +(2\bar{\nu}_{e})\\
(A,Z)\rightarrow(A,Z-2) +2e^{+} +(2{\nu}_{e})\\
e^{-} + (A,Z)\rightarrow(A,Z-2) + e^{+} + (2{\nu}_{e})\\
2e^{-} + (A,Z)\rightarrow(A,Z-2) + (2{\nu}_{e})
\end{eqnarray}

Triggered by the important implication of neutrino mass, a new generation of
experiments are aimed to observe $0\nu\beta\beta$ decay in various isotopes and
with different experimental techniques \cite{dbd-expt-rev}. Among the main 
experimental activities, we remind that recently three
$\beta\beta$ experiments have published new experimental data: GERDA \cite{gerda}, EXO-200 \cite{exo} and 
KamLAND-Zen \cite{kam}.
GERDA is searching for $0\nu\beta\beta$ decay in $^{76}$Ge while EXO-200 and KamLAND-Zen
are looking for the decay in $^{136}$Xe. Several other experiments are going through their
R\&D phase. These include CUORE ($^{130}$Te) \cite{cuore}, SuperNEMO ($^{82}$Se) \cite{supernemo}, 
MAJORANA ($^{76}$Ge) \cite{majorana},
SNO+ ($^{130}$Te) \cite{sno+}, NEXT ($^{136}$Xe) \cite{next} etc.. Present experiments are sensitive to
half-lives of the order of $10^{25}$ years and a corresponding effective mass of the neutrino
$\langle m_{\beta\beta} \rangle \sim 100$ meV \cite{mass}. Constrained by the uncertainties in the
calculation of nuclear transition matrix elements \cite{rath}, phase space factor \cite{kotila} and by the $g_{A}$ value \cite{vis14} that
are used to determine $\langle m_{\beta\beta} \rangle$ as well as by the different background for different isotopes,
it is essential to measure the half-life for  $0\nu\beta\beta$ decay in several isotopes.

Although Sn was considered as one of the potential candidates since the 1980s \cite{doi85},
there have been only a few studies on Sn. It can be mentioned that one of the earliest
attempts for the experimental study of double beta decay was done by Fireman \cite{fire}
in 1949 using $^{124}$Sn.

Natural tin  contains three  isotopes which can decay via double beta  transition;
$^{122,124}$Sn through two electron  mode and $^{112}$Sn through $\beta^{+}$EC and
EC-EC processes, as given in the following:
\begin{eqnarray}
^{122}\text{Sn} \rightarrow {^{122}}\text{Te} + 2e^{-} + (2\bar{\nu}_{e})\\
^{124}\text{Sn} \rightarrow {^{124}}\text{Te} + 2e^{-} + (2\bar{\nu}_{e})\\
e^{-} + {^{112}}\text{Sn} \rightarrow {^{112}}\text{Cd} + e^{+} +(2\nu_{e})\\
2e^{-} + {^{112}}\text{Sn} \rightarrow {^{112}}\text{Cd} + (2\nu_{e})
\end{eqnarray}
The Q-values for  the decay transitions in $^{112}$Sn, $^{122}$Sn  and $^{124}$Sn
are 1919.82$\pm$0.16 keV, 372.9$\pm$2.7 keV  and 2291.1$\pm$1.5 keV \cite{ame}, respectively.  
The natural isotopic abundances for these isotopes are 0.97(1)$\%$, 4.63(3)\% and 
5.79(5)$\%$ \cite{berg}, respectively. The gamma rays from  $^{122}$Sn decays are at 
low energy where the background is  high and makes it more difficult
to study this decay modes in an external source experiment. Therefore, 
we have not considered this nucleus for the present study.

Recently, investigations were carried out using natural tin samples for
$\beta^{-}\beta^{-}$ decay of $^{124}$Sn to the excited states of the daughter
nucleus $^{124}$Te and $\beta^{+}$EC and EC-EC processes in $^{112}$Sn by
 Dawson \textit{et al.}
~\cite{dawson, dawson1}, Kim \textit{et al.}  ~\cite{kim} and 
Barabash \textit{et al.}~\cite{bash3}. Half-life limits of the order of $10^{18} - 10^{21}$ 
years were obtained in those experiments.  Searches for $\beta^{+}$EC and EC-EC processes in
$^{112}$Sn were also carried out  by Kidd \textit{et
  al.} ~\cite{kidd} and Barabash \textit{et  al.}  ~\cite{bash1, bash2} using enriched material obtaining  the best 
half-life limit of $10^{21}$ years. The Kims collaboration ~\cite{kims} gave a half-life limit 
2.0$\times$10$^{20}$ years for neutrino-less double beta decay of $^{124}$Sn using tin-loaded 
liquid scintillator.

An attempt to study the feasibility of a new experiment to search for double beta decay in  $^{112}$Sn 
and $^{124}$Sn was carried out using an ultra-low background HPGe detector. 
The gamma rays produced by the de-excitations of the excited levels of either $^{112}$Cd or $^{124}$Te can 
be detected by the HPGe detector. The aim of the present study was to investigate the decay 
processes to the ground state  as well as  
to the excited states in $^{112}$Sn and to the excited states in $^{124}$Sn using $\gamma$ -ray 
spectrometry. We will discuss in the last 
section also a possible enhancement of the sensitivity with a  proposed experiment 
with larger mass to reach the half-life values estimated by theory.

\section{Experimental set-up and measurements }

\begin{figure}[h]
\begin{center}
\includegraphics[width=1.0\columnwidth, angle=0]{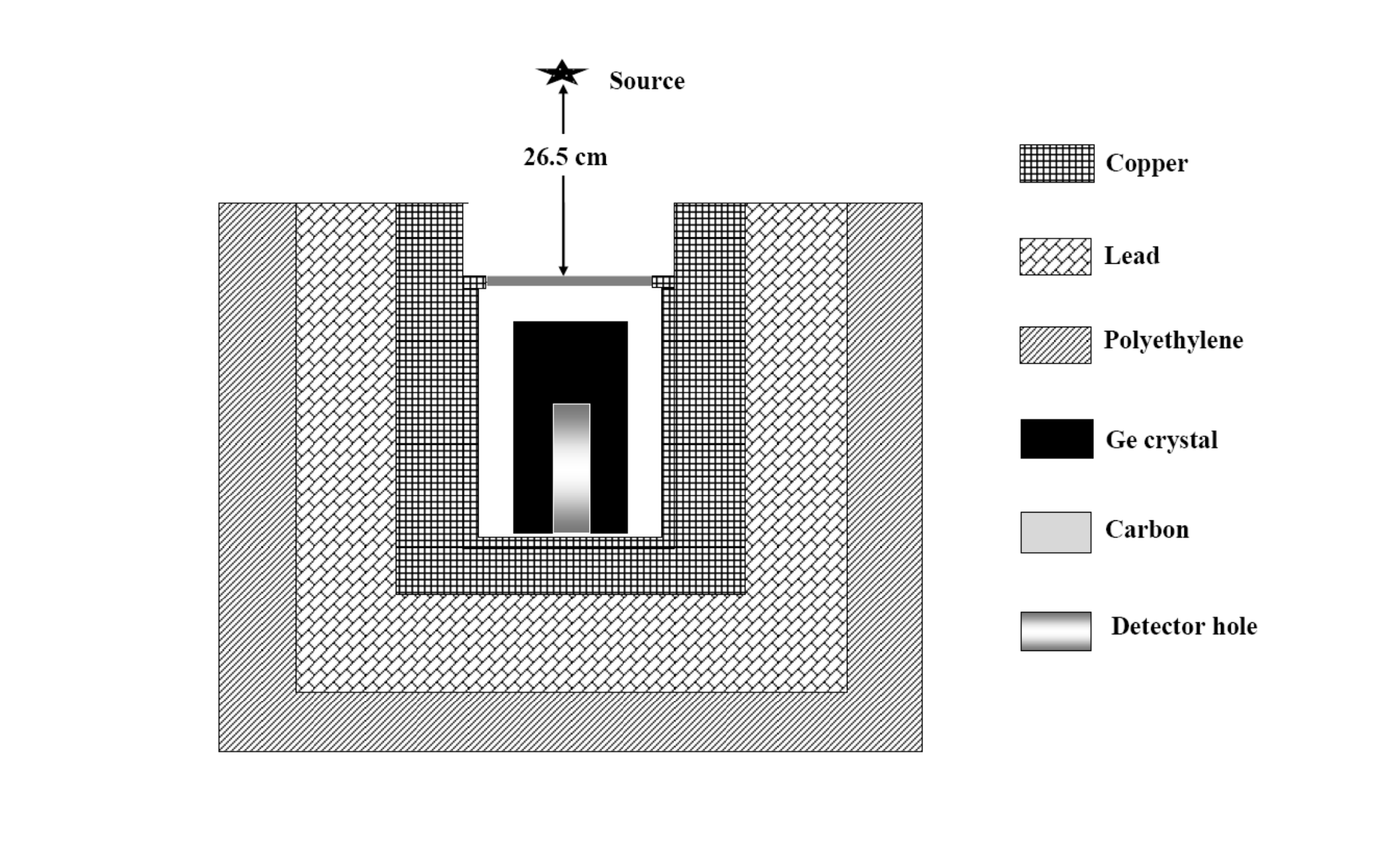}
\vspace{-1cm}
\caption{Schematic diagram of experimental set-up used in simulation. Figure not in scale.}
\label{fig:hpge1}
\end{center}
\end{figure}

\begin{figure}[h]
\begin{center}
\includegraphics[width=1.0\columnwidth, height=6.0 cm]{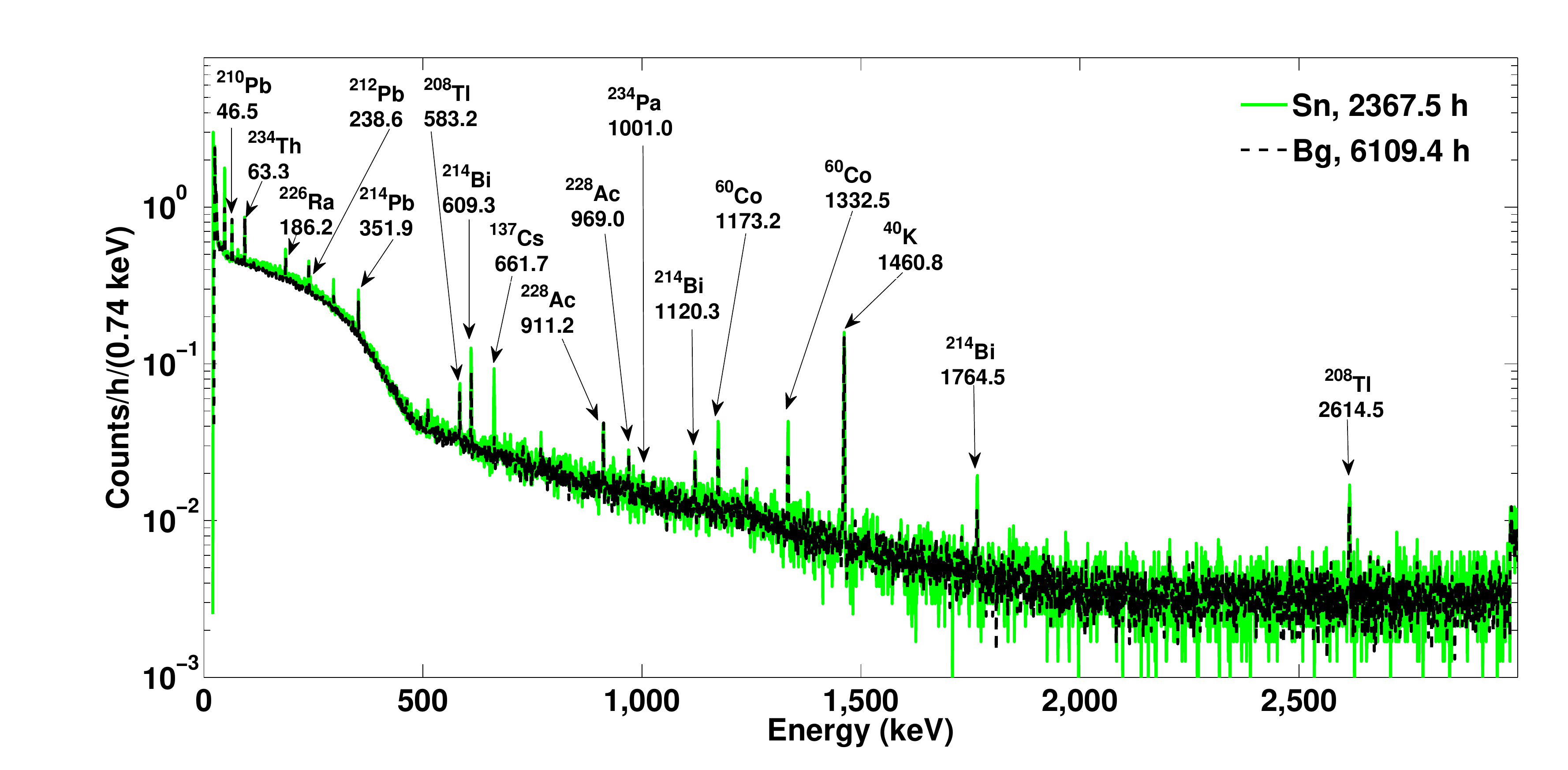}
\caption{Energy spectrum with 13.3 g of natural tin sample (Sn) for 2367.5 h 
of measurement in comparison with background spectrum (Bg) of ultra low-background 
HPGe detector measured for 6109.4 h.  The energy of the $\gamma$ lines are in keV.}
\label{fig:spectra}
\end{center}
\end{figure}  
A  coaxial closed-end n-type ultra low  background HPGe (GeBer) $\gamma$ detector (244 cm$^{3}$) was used 
for the measurement. The detector is placed in the Gran Sasso National Laboratory (LNGS) of the INFN, 
Italy, located underground at $\approx$ 3600 meters  of water equivalent.  The  schematic diagram  
of the experimental  setup is  shown in Fig.~\ref{fig:hpge1}.  The physical dimensions of the detector 
are given in Table~\ref{table:hpge-par}. More details about the detector and the ultra low background setup 
can be found in \cite{proc}. 

Monte-Carlo simulation of HPGe detector has been performed using the GEANT4 software library \cite{geant}. 
To validate the results of the simulation, we have compared the simulated efficiencies with known experimental 
results. The full-energy-peak  (FEP) efficiency of  the HPGe detector was measured for 
different $\gamma$ ray energies using $^{241}$Am, $^{133}$Ba, $^{137}$Cs and $^{60}$Co  point 
sources. The sources were placed at a distance of 26.5 cm above the end cap of the detector. 
The energy resolution (FWHM) of the detector was measured as 2.0 keV at 1332.5 keV. 

A natural tin (purity 99.997$\%$) sample of 13.3 g and with a thickness of 4.5 mm was placed on the end-cap, 
and  the spectrum was measured with the HPGe detector in order to study the double beta decay 
processes in $^{112}$Sn and $^{124}$Sn nuclei. Data were  accumulated
for 2367.5 h  with the sample and for  6109.4 h without the sample (background). The
energy spectra of  sample and background, normalized to the time of measurement of sample, 
are shown  in Fig.~\ref{fig:spectra}. 

\begin{table}[h!]
  \caption[short]{The geometrical parameters of the detector.
It is worth noting that the only difference between optimized and nominal values of the detector is the hole inner dead layer 
(0.3 mm nominal value). See text.}
\begin{center}
\begin{tabular}{l c}
\hline
Detector parameter            &Optimized value (MC2)(mm)\\
\hline
Ge crystal radius                          &30.85\\
Ge crystal length                          &81.70\\
Carbon (C) window thickness                &0.76\\
Ge crystal-C window distance               &4.00 \\
Hole radius                                &4.95\\
Hole length                                &73.40\\
Ge side dead layer                         &0.0\\
Ge front dead layer                        &0.0\\
Hole inner dead layer                      &5.00\\
\hline
\end{tabular}
\end{center}
\label{table:hpge-par}
\end{table}

\section{Results and Discussion}

\subsection{Efficiency of the detector}

\begin{table}[h!]
  \caption[short]{Comparison of experimental efficiency with simulated efficiency 
computed with two different Monte-Carlo codes: EGS4 (E4) and GEANT4 (G4). 
Efficiencies related to MC1 and MC2 are calculated considering two different 
thickness of the inner dead-layer: 1.5 mm and 5.0 mm, respectively.}
\begin{center}
\resizebox{\textwidth}{!}{
\begin{tabular}{l l l l l l l l l l l}
\hline
Source &Energy    &Exp   &MC1 &$\frac{Exp}{MC1}$ &MC2 &$\frac{Exp}{MC2}$  &MC1 &$\frac{Exp}{MC1}$  &MC2     &$\frac{Exp}{MC2}$\\
       &keV       &$\%$  &(E4)$\%$ &(E4)         &(E4) $\%$&(E4)          &(G4)$\%$&(G4)           &(G4)$\%$ &(G4)          \\
\hline
$^{241}$Am&26.3     &0.0048&0.0064   &75$\%$        &0.0064    &75$\%$        &0.0066    &73$\%$       &0.0064   &75$\%$ \\
          &59.5     &0.0938&0.1089   &86$\%$        &0.1088    &86$\%$        &0.1135    &83$\%$       &0.0974   &96$\%$ \\
$^{133}$Ba&81.0+79.6&0.0913&0.1073   &85$\%$        &0.1044    &87$\%$        &0.1033    &88$\%$       &0.0979   &94$\%$ \\
         % &+79.6    &      &         &              &          &         &          &        &         &  \\
          &276.4    &0.0111&0.0131   &85$\%$        &0.0114    &97$\%$        &0.0132    &84$\%$       &0.0116   &96$\%$ \\
          &302.9    &0.0270&0.0305   &89$\%$        &0.0271    &100$\%$       &0.0313    &86$\%$       &0.0280   &96$\%$ \\
          &356.0    &0.0798&0.0926   &86$\%$        &0.0799    &100$\%$       &0.0955    &84$\%$       &0.0810   &99$\%$ \\
          &383.8    &0.0110&0.0126   &87$\%$        &0.0107    &103$\%$       &0.0125    &88$\%$       &0.0109   &101$\%$ \\
$^{137}$Cs&661.7    &0.0668&0.0818   &82$\%$        &0.0679    &98$\%$        &0.0948    &71$\%$       &0.0690   &97$\%$ \\
$^{60}$Co &1173.2   &0.0527&0.0652   &81$\%$        &0.0531    &99$\%$        &0.0680    &78$\%$       &0.0520   &101$\%$ \\
          &1332.5   &0.0478&0.0600   &80$\%$        &0.0476    &100$\%$       &0.0636    &75$\%$       &0.0480   &100$\%$ \\
\hline
\end{tabular}
}
\end{center}
\label{table:efficiency}
\end{table}
The FEP efficiency of the detector was  simulated for the calibrated point sources using
Monte-Carlo based software libraries  GEANT4 ~\cite{geant} and EGS4 ~\cite{egs}. Generally, the 
results of the  Monte Carlo simulations deviate significantly ($> 10\%$) from
the experimental results ~\cite{osaka, jhon, chham,huy, azli, cabal, hur}. The difference can be due to two 
reasons: the uncertainties associated with the detector shape parameters
provided by the manufacturer or incomplete charge collection in the
crystal during the measurement process. 
\begin{figure}[h]
\begin{center}
\includegraphics[width=1.0\columnwidth, angle=0]{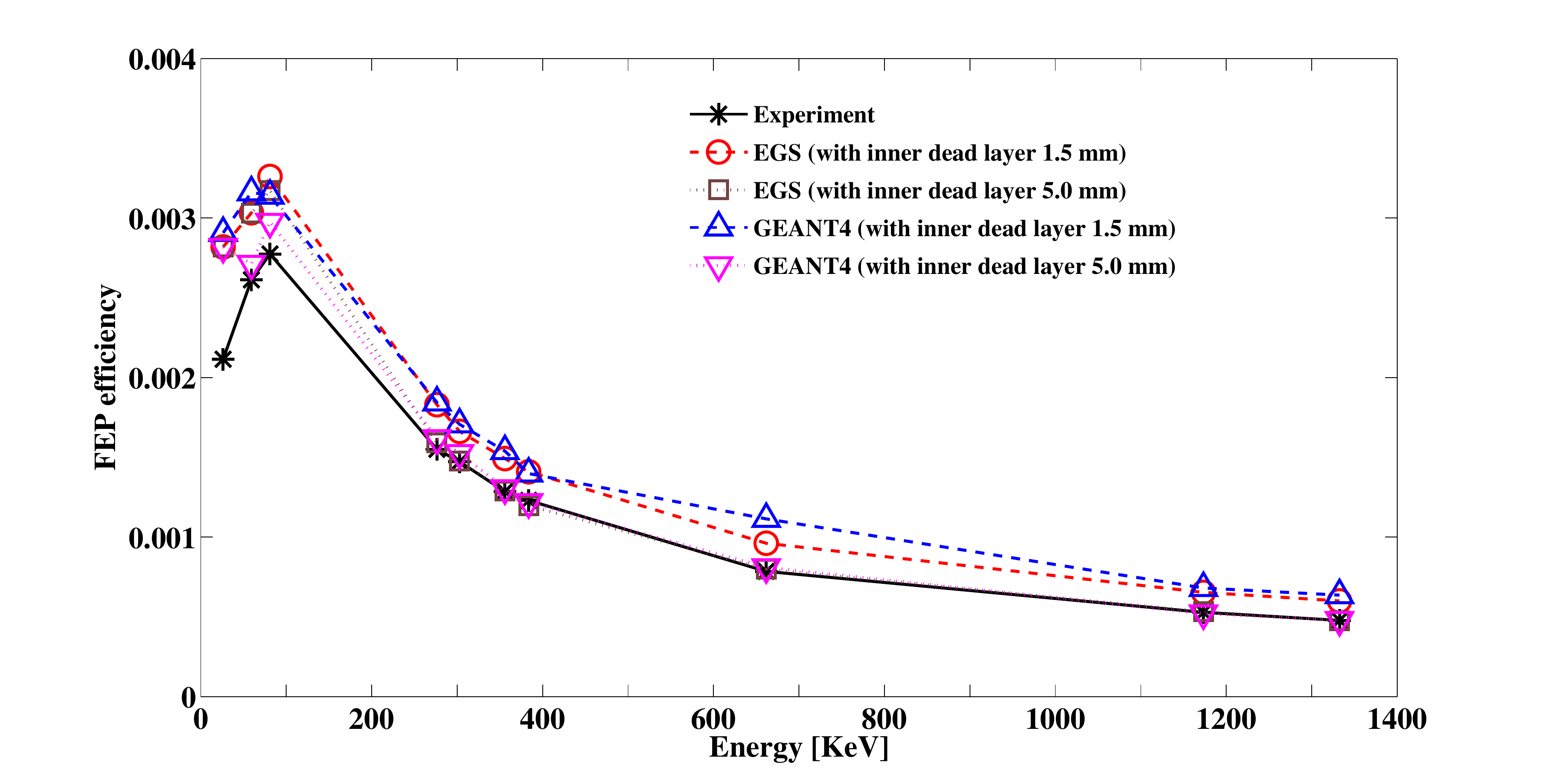}
\caption{Comparison of the experimental efficiencies with those obtained from Monte-Carlo 
simulation using two different codes EGS4 and GEANT4. Here the FEP efficiencies have been 
divided by the branching ratios of the corresponding gammas.}
\label{fig:effi-comp}
\end{center}
\end{figure} 
Usually, all the information about the inner
components of the germanium detector are not supplied by the manufacturer. Thus uncertainties arise
due to possible additional absorbers or on the nature of
the materials allocated around the Ge crystal.
Furthermore, the dimensions provided by the manufacturer correspond
to the time of assembly of the detection system at room temperature.
Then changes in the mechanical support of the crystal due
to contractions at low temperature may lead to changes in the detector
configurations ~\cite{jhon}.
There is also a certain uncertainty in the crystal parameters e.g.
the dead layer thicknesses ~\cite{chham} or the distance between end cap and
Ge crystal.
It has also been observed, that there can be a substantial increase
of the dead layer thickness after some years of operating time ~\cite{huy}.
Therefore, to match the
experimental  efficiency values, the detector parameters and the
dead layer thickness can be varied ~\cite{hur}.  In the first step, the
detector parameters provided by the manufacturer was used in the simulation.  
A fine adjustment was then made by varying only the inner dead
layer thickness  in a systematic way to match the experimental efficiency values. 
In order to show the dependence of the calculated efficiencies on the inner 
hole dead layer, the results obtained by considering two different thicknesses 
(1.5 mm and 5.0 mm) are compared with the experimental values in Table~\ref{table:efficiency} and 
in Fig.~\ref{fig:effi-comp}. In the latter case the FEP efficiencies have been divided by the 
branching ratios of the corresponding $\gamma$ rays. From Table~\ref{table:efficiency}, we can see 
that for the optimized dead layer of the inner hole, thickness of 5.0 mm, the simulated 
efficiencies (for both EGS4 and GEANT4) are in very good  
agreement with the experimental results except for the very low energy region. The uncertainty is 
within 4$\%$, which is quite reasonable considering the  statistical fluctuations. 
However, at very low energy, difference is quite large. This could be due to the
very small mean free path of $\gamma$ rays at this energy scale. Therefore, 
some $\gamma$ rays are absorbed before reaching  the detector. 
 The optimized parameters were then used to calculate the FEP
efficiency for a volume source of dimensions equal to the tin sample used in
the present work.

\subsection{Radioactive contamination in the sample}

The sources of background radiations measured in an underground laboratory can
be classified into several categories; environmental radioactivity including
radon, gamma rays, neutrons from
natural fission and from the ($\alpha$,n) reaction, radioactive impurities in
the detector and shielding materials and cosmic rays with relevant
contributions from muons and neutrons \cite{boudis,klap}. In the measured spectra, 
background components can be identified by their characteristic $\gamma$-emission 
peaks.  
\begin{table}[h!]
\caption[short]{\label{cont} The radioactive contamination present in the Tin sample
measured with HPGe detector. Limits are given at 90 $\%$ C. L.}
\begin{center}
\begin{tabular}{ c c c c c c}
\hline
Source   & Energy & Count rate (mBq)  & Count rate (mBq)    &$\eta$  & Activity in\\
         & (keV)  & from Signal      & from background    &($\%$)  &(mBq/Kg)\\
\hline
$^{228}$Ac& 911.2  &0.014$\pm$0.002     &0.012$\pm$0.001    &1.16    &$\le30$ \\
$^{212}$Pb& 238.6  &0.053$\pm$0.005     &0.048$\pm$0.003    &6.94    &$\le15$\\
$^{208}$Tl& 583.2  &0.021$\pm$0.002     &0.015$\pm$0.001    &5.83    &7$\pm$3\\
$^{212}$Bi& 727.3  &0.004$\pm$0.001     &0.006$\pm$0.001    &0.37    &$\le34$\\
$^{214}$Bi& 1764.5 &0.098$\pm$0.001     &0.006$\pm$0.001   &0.41    &42$\pm$22  \\
$^{214}$Pb& 351.9  &0.048$\pm$0.004     &0.033$\pm$0.002    &3.81    &30$\pm$9\\
$^{210}$Pb&  46.5  &0.335$\pm$0.008     &0.196$\pm$0.004    &1.30    &801$\pm$51\\
$^{234}$Th&  63.3  &0.095$\pm$0.005     &0.103$\pm$0.004    &1.11    &$\le 28$\\
$^{226}$Ra&  186.2 &0.056$\pm$0.006     &0.060$\pm$0.004    &0.72    &$\le 79$ \\
$^{234m}$Pa&1001.0 &0.004$\pm$0.001     &0.003$\pm$0.001   &0.03    &$\le739$ \\  
$^{137}$Cs& 661.7  &0.030$\pm$0.002     &0.003$\pm$0.001   &5.03    &39$\pm$4\\
$^{60}$Co & 1173.2 &0.020$\pm$0.002     &0.012$\pm$0.001   &3.63    &17$\pm$4  \\
$^{40}$K  &1460.8  &0.094$\pm$0.003     &0.087$\pm$0.002    &0.34    &155$\pm$79 \\
\hline
\end{tabular}
\end{center}
\label{table:activity}
\end{table}
The specific activities of the radioactive nuclei present in
the natural tin sample were calculated with the formula ~\cite{proc}
\begin{equation}
A= \frac{(S_{s}/t_{s}-S_{b}/t_{b})}{m \eta }, \label{activity}
\end{equation}
where $S_{s}$ and $S_{b}$ denote the area under a peak in the sample and
background spectra, respectively.  The measurement time of the sample and
background spectra are denoted by $t_{s}$ and $t_{b}$, respectively. The mass of
the sample is represented by $m$ and $\eta$ is the efficiency of the full-energy
peak detection ($\eta$ also takes into account the decay fraction of the $\gamma$ ray which was
obtained from National Nuclear Data Center (NNDC) ~\cite{nndc}). Activities of the natural 
radioactive sources 
present in the tin sample are listed in Table~\ref{table:activity}. Activity limits are given 
at 90 $\%$ confidence level (C. L.) according to the Feldman-Cousins method \cite{feld}.

\subsection{ Double beta decay study of $^{112,124}$Sn}

In the accumulated spectra with the tin sample, no peaks are observed which could be 
unambiguously attributed to the double beta decay processes of $^{112,124}$Sn. Therefore, 
only half-life limits are  calculated using the formula
\begin{equation}
T_{1/2} \ge \frac{(\text{ln}~2)\cdot N \cdot \eta\cdot t}{\text{lim}~S},
\end{equation}
where $N$  is the number  of $\beta\beta$-active  nuclei, $t$ is  the measurement
time and $\eta$ is  the FEP detection efficiency.  The expression lim~$S$ is the number of events of the
effect searched for which can be excluded at a given C. L.. All the limits reported 
in the present study  are given at 90$\%$ C. L.. The values of lim~$S$  were calculated using the 
Feldman-Cousins 
procedure \cite{feld}. The detection efficiencies of the double beta processes in the tin isotopes were 
calculated using  the GEANT4 software library ~\cite{geant}.
The DECAY0 \cite{decay} event generator has been used to generate the initial kinematics of the particles.

Taking into account  the isotopic composition of natural tin, the sample
contained  $6.55\times10^{20}$ nuclei of $^{112}$Sn
and $3.90\times10^{21}$ nuclei of $^{124}$Sn. The measured limits on half-lives for these 
two isotopes along with today's best experimental limits and theoretical estimates are reported in 
Table~\ref{tab:thalf-limit}. 

\begin{figure}[!ht]
\begin{center}
\includegraphics[width=1.0\columnwidth, angle=0]{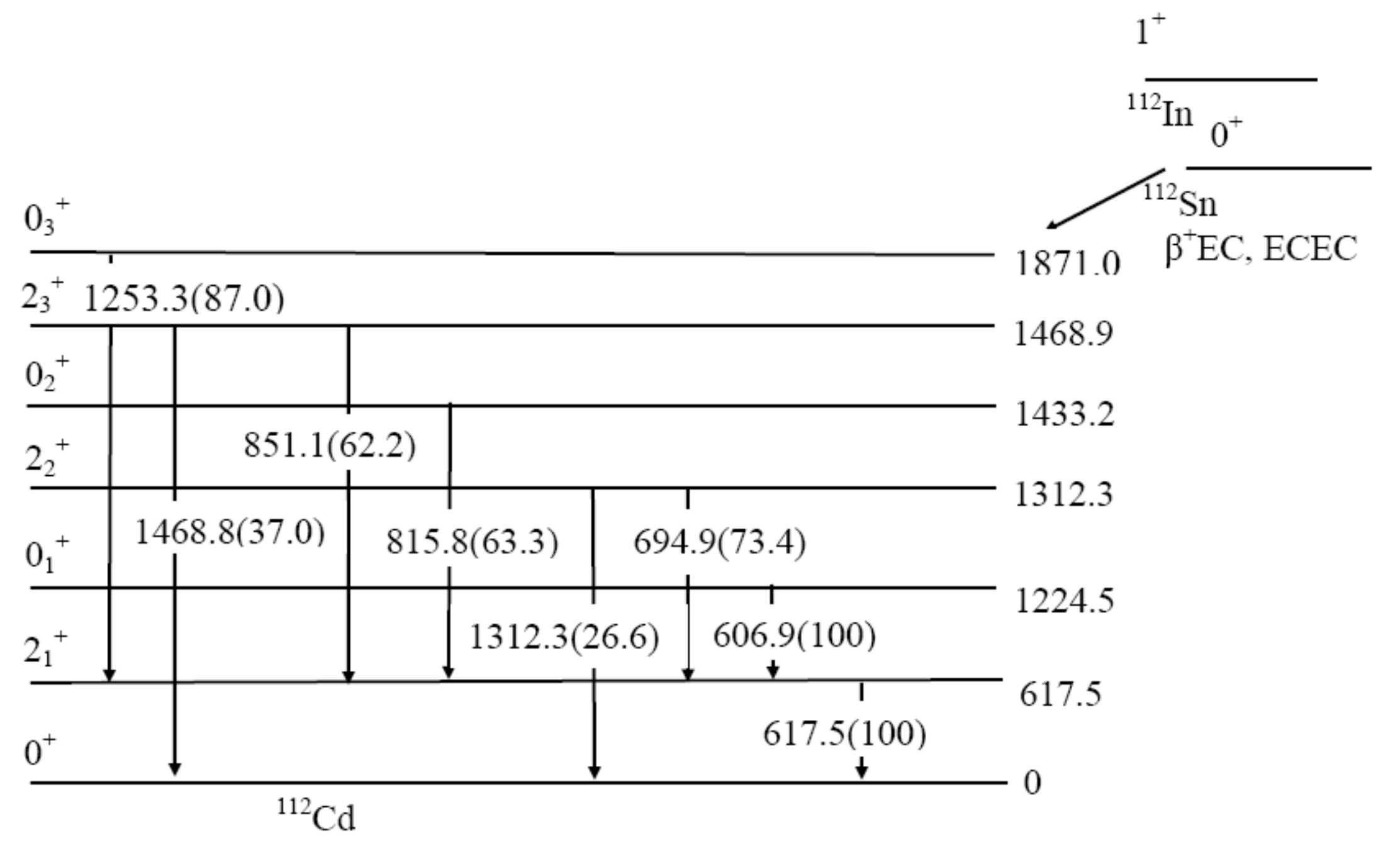}
\vspace{-1cm}
\caption{Partial decay scheme of $^{112}$Sn \cite{nds96}. The energies of the excited 
levels and the emitted $\gamma$ quanta 
are in keV with relative intensities of $\gamma$ quanta  are given in parenthesis. }
\label{fig:112sn-scheme}
\end{center}
\end{figure}

\subsubsection{EC-EC decay of $^{112}$Sn}

In the 2$\nu$ EC-EC decay of $^{112}$Sn to the ground state of $^{112}$Cd, all the energy release
 is carried away by the neutrinos except for a very small amount emitted as X-rays. These X-rays 
lie below the energy threshold of the measurement apparatus. In case of the neutrino-less mode, 
bremsstrahlung $\gamma$ quanta are emitted with an energy equal to 
$E_{\gamma} = Q_{2\beta} - \epsilon_{1} -\epsilon_{2} - E_{exe},$
where $\epsilon_{i}$ are electron binding energies of daughter nuclide, and E$_{exe}$ is the 
populated level energy of $^{112}$Cd. The partial decay scheme of $^{112}$Sn is shown in 
Fig.~\ref{fig:112sn-scheme} \cite{nds96}. For the transition to the excited state, the bremsstrahlung $\gamma$ quanta 
are accompanied by $\gamma$ rays emitted from nuclear de-excitation. 
We did not observe any peak with the expected energies for the EC-EC decay of  $^{112}$Sn. Only lower 
limits of half-lives are obtained using the Feldman-Cousins prescription \cite{feld}. 
The T$_{1/2}$ limits along with the $\gamma$ energies and the corresponding detection 
efficiencies are shown in Table~\ref{tab:thalf-limit}. 

\begin{figure}[h!]
\begin{center}
\includegraphics[width=0.8\columnwidth, angle=0]{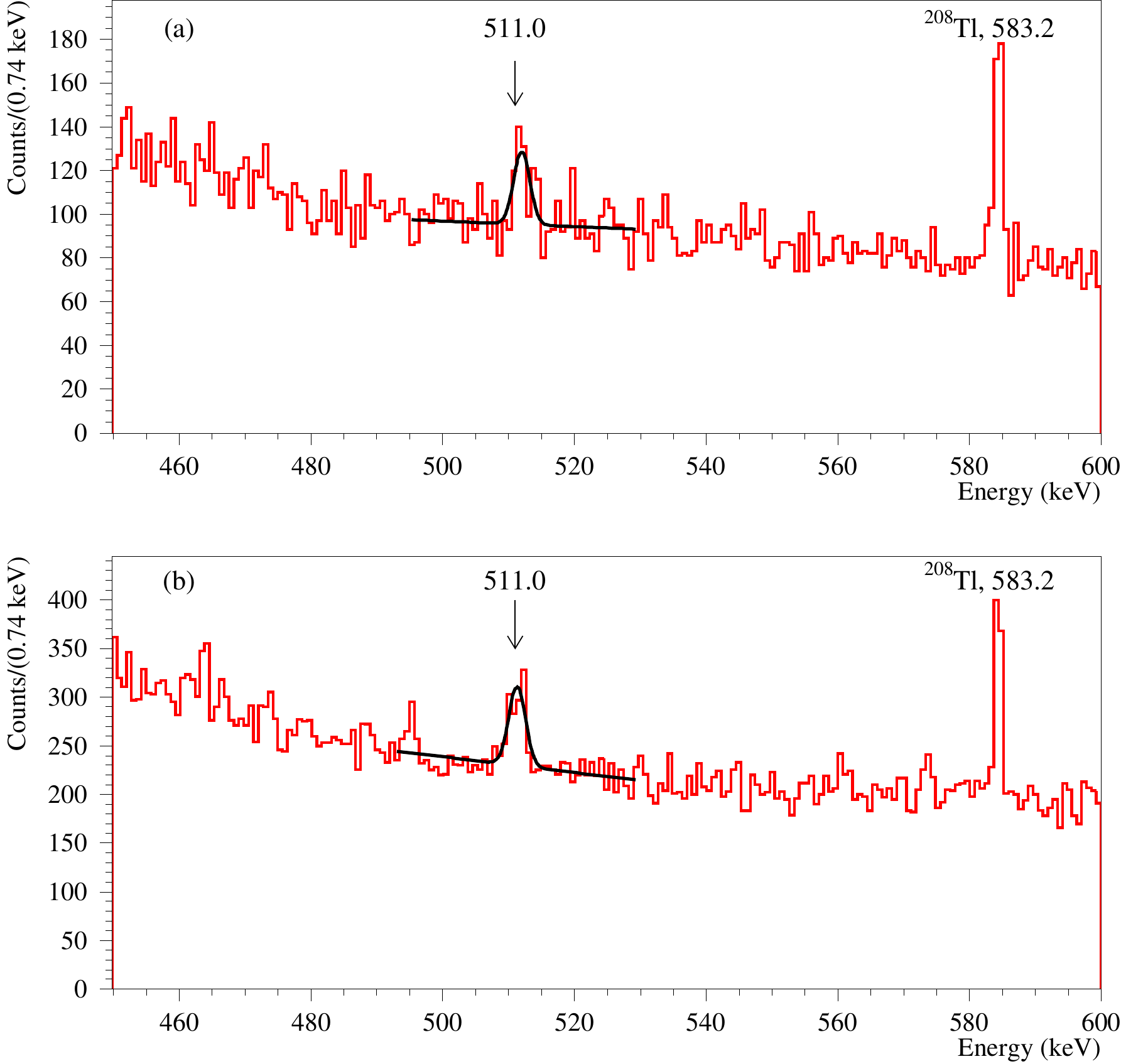}
\caption{Part of the energy spectra accumulated with Sn sample for 2367.5 h of measurement 
by ultralow-background HPGe $\gamma$ detector for energy range (450-600) keV (upper panel). 
The energy spectrum accumulated without the sample over 6109.4 h is also shown (lower panel). 
Fits of the 511 keV annihilation  $\gamma$  peaks are shown by solid line.} 
\label{fig:511fit}
\end{center}
\end{figure}

\subsubsection{$\beta^{+}$EC processes in $^{112}$Sn}

The ($0\nu + 2\nu)\beta^{+}$EC transition to the ground state of daughter nuclei is accompanied by two
 annihilation $\gamma$ quanta with  energy 511.0 keV. Moreover, the detector surroundings and the sample 
contains the radioactive element  $^{208}$Tl
which emits $\gamma$ quanta of energy 510.8 keV. The energy spectra accumulated with and without the 
sample in the energy interval 450-600 keV are presented in Fig.~\ref{fig:511fit}.
The measured area of the annihilation peak is (130$\pm$7) counts. The area of the peak in the 
background spectrum is (100$\pm$15) counts (normalized on the time of measurement with the tin sample). 
The excess of events in the data accumulated with the tin sample can be explained by radioactive contamination
of the sample itself. In particular, the decay of $^{208}$Tl contributes 17$\pm$7 counts of 510.8 keV gamma quanta.
The difference in the areas of the annihilation peak (13$\pm$18) counts, which can be attributed to 
electron capture with positron emission in $^{112}$Sn, gives no indication on the effect.
In accordance with the Feldman-Cousins procedure, 42.6 counts can be excluded
at 90\% C.L.
Taking into account the FEP efficiency ($\eta$) of  7.58$\%$, the lower limit of the 
($0\nu + 2\nu)\beta^{+}$EC transition to the ground state is 2.18$\times 10^{17}$ yr.

In the case of the ($0\nu + 2\nu)\beta^{+}$EC transition to the excited 2$_{1}^{+}$ (617.5 keV) states 
of $^{112}$Cd, no peak has been observed. The limit on the half-life is reported in Table~\ref{tab:thalf-limit}. 
The theoretical  T$_{1/2}$ is estimated to be very high because of the low phase space available for 
this transition.

\subsubsection{The $\beta ^- \beta ^-$ decay of $^{124}$Sn}

\begin{figure}[h!]
\includegraphics[width=1.0\columnwidth, angle=0]{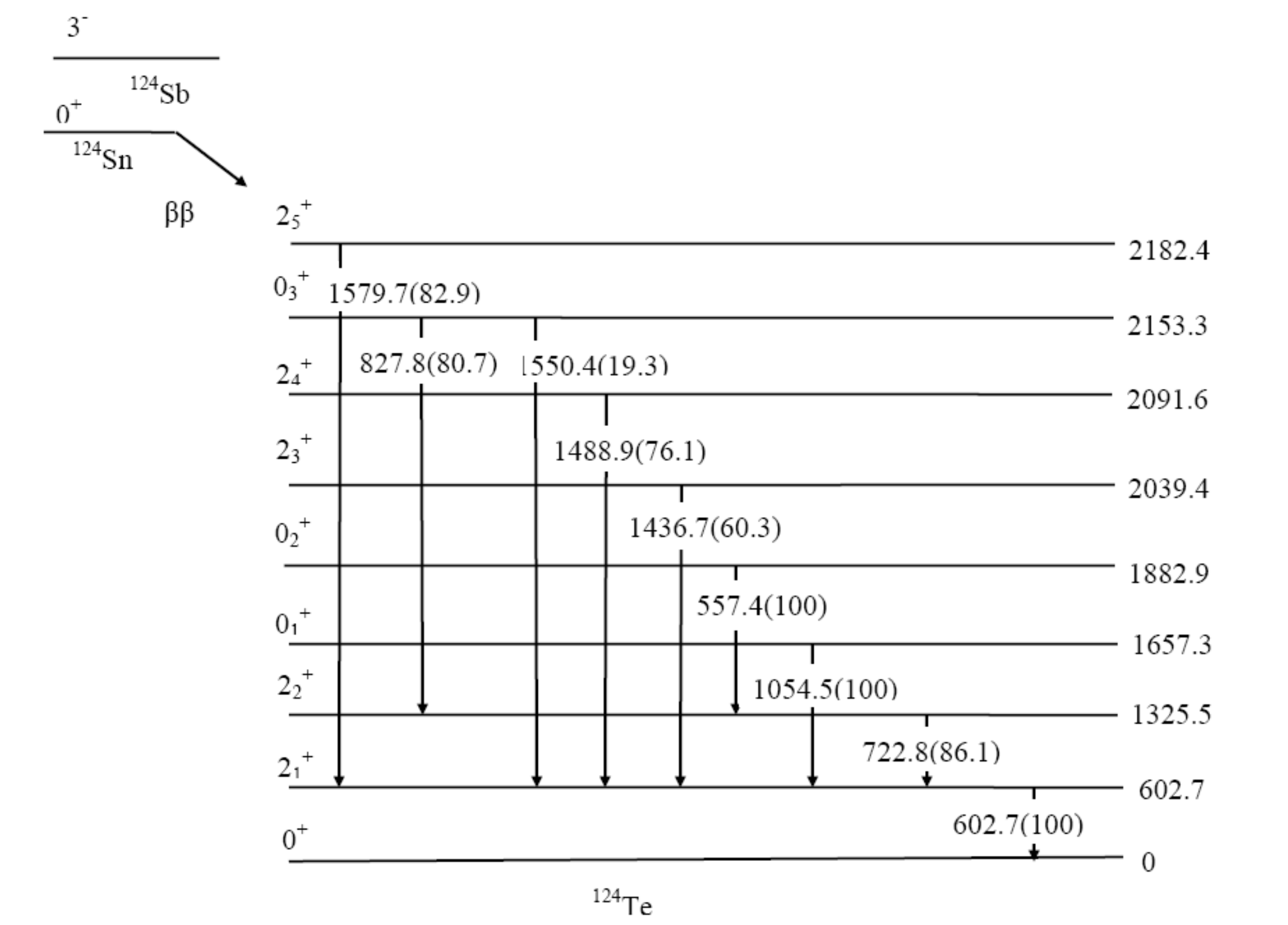}
\caption{Partial decay scheme of $^{124}$Sn \cite{nds08}. The energies of the excited levels and of 
the $\gamma$ quanta are in keV with relative intensities 
of $\gamma$ quanta are given in parenthesis.}
\label{fig:124Sn-scheme}
\end{figure}

\begin{figure}[h!]
\begin{center}
\includegraphics[width=0.8\columnwidth ]{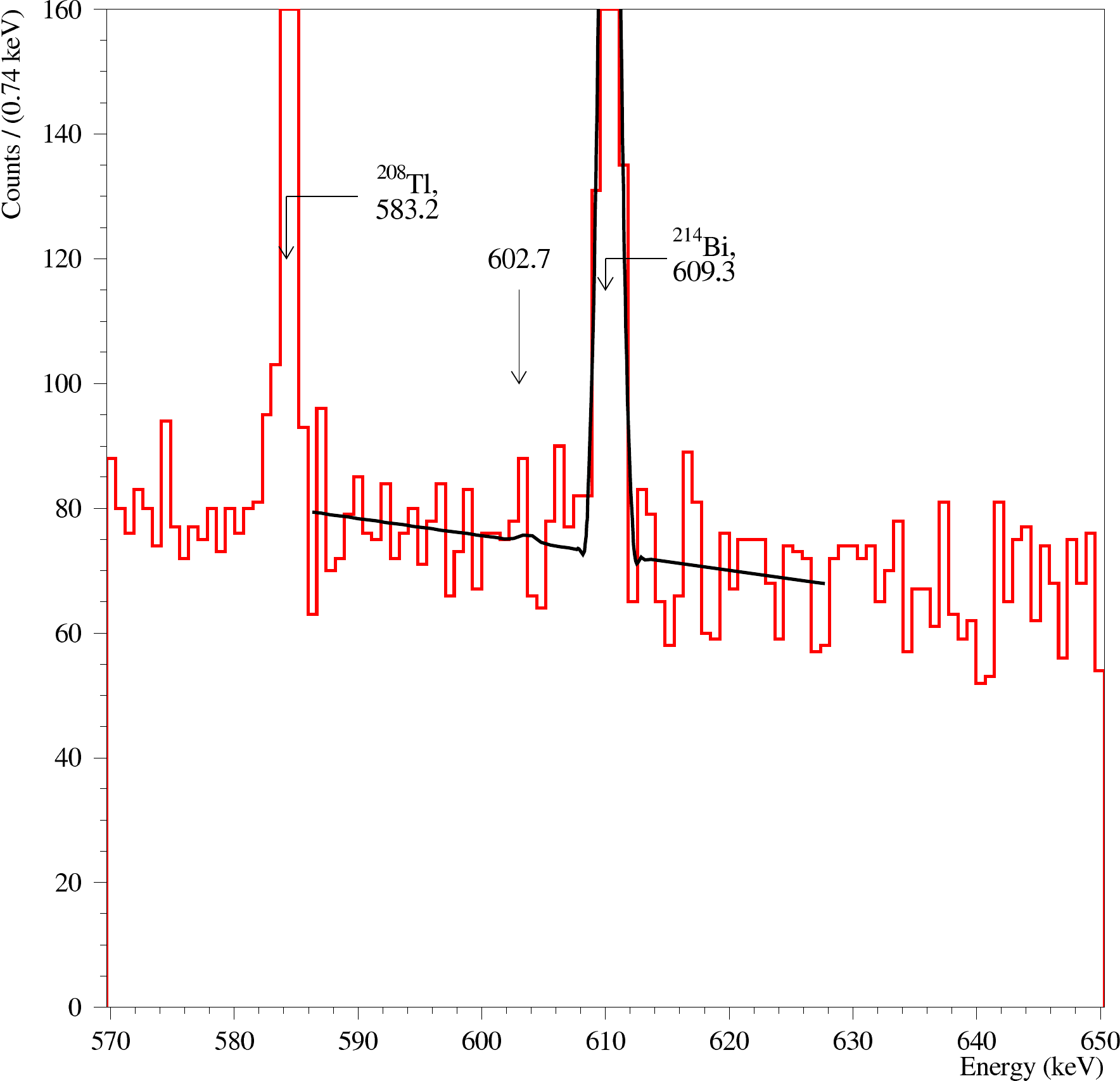}
\caption{Part of the energy spectra accumulated with Sn sample for 2367.5 h of measurement. 
In particular, the region around the 602.7 keV peak, expected for the $\beta^{-}\beta^{-}$ processes in $^{124}$Sn to the
excited states of  $^{124}$Te, is shown. Fits of the excluded peak at energy 602.7 keV along with 
peak due to $^{214}$Bi with energy 609.3 keV are depicted. Peak of $^{208}$Tl with 
energy 583.2 keV is also shown.}
\label{fig:124sn-ecec}
\end{center}
\end{figure}

For $\beta ^- \beta ^-$ decay of $^{124}$Sn, only transitions to excited states of $^{124}$Te can be 
studied as we are measuring gamma quanta. The partial decay scheme of $^{124}$Sn is shown in 
Fig.~\ref{fig:124Sn-scheme} \cite{nds08}.  No peak has been observed at  602.7 keV, as shown in Fig.~\ref{fig:124sn-ecec}. 

To derive the lim~$S$ value for the 602.7 keV $\gamma$- line, a model of the background was built from
two Gaussian (one to describe the effect, and the second one to take into account the gamma
peak with energy 609.3 keV of $^{214}$Bi) plus first-degree polynomial to describe the continuous background.
The best fit ($\chi^{2}$/n.d.f. = 50.8/51 = 0.99) achieved
in the energy interval 586--628 keV gives the area of the peak searched for 2$\pm$12 counts
(at the energy 602.7 keV), which conservatively provides lim$~S$ = 22 counts.
Considering the transition to the 2$_{5}^{+}$ (2182.4 keV) state (studied here for the first time), the calculated 
efficiency to detect $\gamma$ quanta with energy 602.7 keV is 4.06\%, thus T$_{1/2}$ $\ge$ 1.36$\times 10^{18}$ yr.
The half-lives limits for the transition of $^{124}$Sn to other excited states of
$^{124}$Te are reported in the lower part of Table~\ref{tab:thalf-limit}.

The overall result is that due to the small mass of the tin sample ( 13.3 g), the obtained limits 
are rather poor compared to the existing experimental data. Therefore, it is certainly worthwhile to 
repeat  the measurement with a larger sample, may be even enriched in the isotopes of interest, 
in order to obtain more stringent limits. 
\begin{table*}[!htbp]
\renewcommand{\arraystretch}{0.7}
\setlength{\extrarowheight}{-2.0pt}
\caption[short]{The experimental limits  and theoretical predictions for $\beta^{+}$EC/ECEC decay in $^{112}$Sn and 
$\beta^{-}\beta^{-}$ decay in $^{124}$Sn.  }
\begin{center}
\resizebox{\textwidth}{!}{
\begin{tabular}{lccccccc}
\hline
Transition & Mode & Energy of & $\eta $ & $LimS$ & \multicolumn{2}{c}{$%
T_{1/2}^{exp}$(yr) at 90$\%$ C. L.} & $T_{1/2}^{th}(2\nu )(yr)$ \\ 
\cline{6-7}
&  & $\gamma $ rays & ($\%)$ & 90$\%$ C. L & Present & Previous & \cite
{domi, suho2} \\ 
&  & (keV ) &  &  & work & work \cite{bash3,bash2} &  \\ \hline
\multicolumn{8}{l}{\textbf{$^{112}$Sn $\rightarrow $ $^{112}$Cd}} \\ 
$\beta ^{+}$EC; g.s. & 0$\nu $ + 2$\nu $ & 511.0 & 7.58 & 42.6 & 2.18$\times
10^{17}$ & 0.92$\times 10^{20}$ & 3.8$\times 10^{24}$ \\ 
$\beta ^{+}$EC; $2_{1}^{+}$ 617.5 & 0$\nu $ + 2$\nu $ & 617.5 & 5.95 & 7.8 & 
9.40$\times 10^{17}$ & 7.02$\times 10^{20}$ & 2.3$\times 10^{32}$ \\ 
ECEC; $K^{1}L^{2}$;g.s. & 0$\nu $ & 1889.1 & 2.16 & 2.1 & 12.68$\times
10^{17}$ & 8.15$\times 10^{20}$ & -- \\ 
ECEC; $2_{1}^{+}$ 617.5 & 0$\nu $ & 617.5 & 4.59 & 7.8 & 7.25$\times 10^{17}$
& 9.40$\times 10^{20}$ & -- \\ 
ECEC; $0_{1}^{+}$ 1224.4 & 0$\nu $ & 606.9 & 3.32 & 40.2 & 1.02$\times
10^{17}$ & 12.86$\times 10^{20}$ & -- \\ 
&  & 617.5 & 3.22 & 7.8 & 5.09$\times 10^{17}$ & -- & -- \\ 
ECEC; $2_{2}^{+}$ 1312.3 & 0$\nu $ & 617.5 & 2.35 & 7.8 & 3.71$\times 10^{17}
$ & 8.89$\times 10^{20}$ & -- \\ 
&  & 694.7 & 2.11 & 35.9 & 0.72$\times 10^{17}$ & -- & -- \\ 
&  & 1312.3 & 0.86 & 18.0 & 0.59$\times 10^{17}$ & -- & -- \\ 
ECEC; $0_{2}^{+}$ 1433.2 & 0$\nu $ & 617.5 & 2.85 & 7.8 & 4.50$\times 10^{17}
$ & 6.86$\times 10^{20}$ & -- \\ 
&  & 815.8 & 1.01 & 32.0 & 0.39$\times 10^{17}$ & -- & -- \\ 
ECEC; $2_{3}^{+}$ 1468.8 & 0$\nu $ & 617.5 & 2.06 & 7.8 & 3.26$\times 10^{17}
$ & 6.46$\times 10^{20}$ & -- \\ 
&  & 851.1 & 1.59 & 6.0 & 3.27$\times 10^{17}$ & -- & -- \\ 
&  & 1468.8 & 1.05 & 5.2 & 2.49$\times 10^{17}$ & -- & -- \\ 
ECEC; $0_{3}^{+}$ 1871.14 & 0$\nu $ & 617.5 & 4.76 & 7.8 & 7.52$\times
10^{17}$ & 13.43$\times 10^{20}$ & -- \\ 
&  & 1253.4 & 2.72 & 24.4 & 1.37$\times 10^{17}$ & -- & -- \\ 
ECEC; $2_{1}^{+}$ 617.5 & 2$\nu $ & 617.5 & 6.06 & 7.8 & 9.58$\times 10^{17}
$ & 11.94$\times 10^{20}$ & 4.9$\times 10^{28}$ \\ 
ECEC; $0_{1}^{+}$ 1224.4 & 2$\nu $ & 606.9 & 4.26 & 40.2 & 1.30$\times
10^{17}$ & 16.25$\times 10^{20}$ & 7.4$\times 10^{24}$ \\ 
&  & 617.5 & 4.51 & 7.8 & 7.13$\times 10^{17}$ & -- & -- \\ 
ECEC; $2_{2}^{+}$ 1312.3 & 2$\nu $ & 617.5 & 3.26 & 7.8 & 5.15$\times 10^{17}
$ & 11.24$\times 10^{20}$ & 1.9$\times 10^{32}$ \\ 
&  & 694.7 & 2.96 & 35.9 & 1.01$\times 10^{17}$ & -- & -- \\ 
&  & 1312.3 & 1.25 & 18.0 & 0.85$\times 10^{17}$ & -- & -- \\ 
ECEC; $0_{2}^{+}$ 1433.2 & 2$\nu $ & 617.5 & 3.90 & 7.8 & 6.16$\times 10^{17}
$ & 8.64$\times 10^{20}$ & -- \\ 
&  & 815.8 & 1.39 & 32.0 & 0.54$\times 10^{17}$ & -- & -- \\ 
ECEC; $2_{3}^{+}$ 1468.8 & 2$\nu $ & 617.5 & 2.80 & 7.8 & 4.42$\times 10^{17}
$ & 8.19$\times 10^{20}$ & 6.2$\times 10^{31}$ \\ 
&  & 851.1 & 2.15 & 6.0 & 4.42$\times 10^{17}$ & -- & -- \\ 
&  & 1468.8 & 1.42 & 5.2 & 3.37$\times 10^{17}$ & -- & -- \\ 
ECEC; $0_{3}^{+}$ 1871.14 & 2$\nu $ & 617.5 & 4.76 & 7.8 & 7.52$\times
10^{17}$ & 13.43$\times 10^{20}$ & 5.4$\times 10^{34}$ \\ 
&  & 1253.4 & 2.72 & 24.4 & 1.37$\times 10^{17}$ & -- & -- \\ 
\multicolumn{8}{l}{\textbf{$^{124}$Sn $\rightarrow $ $^{124}$Te}} \\ 
$\beta ^{-}\beta ^{-}$; $2_{1}^{+}$ 602.7 & 0$\nu $+2$\nu $+0$\nu \chi ^{0}$
& 602.7 & 6.05 & 22.0 & 2.02$\times 10^{18}$ & 9.1$\times 10^{20}$ & 4.8$%
\times 10^{23}$ \\ 
$\beta ^{-}\beta ^{-}$; $2_{2}^{+}$ 1325.5 & 0$\nu $+2$\nu $+0$\nu \chi ^{0}$
& 602.7 & 3.94 & 22.0 & 1.32$\times 10^{18}$ & 9.4$\times 10^{20}$ & 2.5$%
\times 10^{27}$ \\ 
&  & 722.8 & 3.40 & 28.0 & 0.89$\times 10^{18}$ & -- & -- \\ 
$\beta ^{-}\beta ^{-}$; $0_{1}^{+}$ 1657.3 & 0$\nu $+2$\nu $+0$\nu \chi ^{0}$
& 602.7 & 4.59 & 22.0 & 1.54$\times 10^{18}$ & 12.0$\times 10^{20}$ & -- \\ 
&  & 1054.5 & 3.21 & 2.7 & 8.71$\times 10^{18}$ & -- & -- \\ 
$\beta ^{-}\beta ^{-}$; $0_{2}^{+}$ 1882.9 & 0$\nu $+2$\nu $+0$\nu \chi ^{0}$
& 557.4 & 3.70 & 7.6 & 3.57$\times 10^{18}$ & 12.0$\times 10^{20}$ & -- \\ 
&  & 602.7 & 2.86 & 22.0 & 0.96$\times 10^{18}$ & -- & -- \\ 
&  & 722.8 & 2.51 & 28.0 & 0.66$\times 10^{18}$ & -- & -- \\ 
$\beta ^{-}\beta ^{-}$; $2_{3}^{+}$ 2039.4 & 0$\nu $+2$\nu $+0$\nu \chi ^{0}$
& 602.7 & 3.07 & 22.0 & 1.03$\times 10^{18}$ & 8.6$\times 10^{20}$ & -- \\ 
&  & 1436.7 & 1.57 & 2.3 & 5.00$\times 10^{18}$ & -- & -- \\ 
$\beta ^{-}\beta ^{-}$; $2_{4}^{+}$ 2091.6 & 0$\nu $+2$\nu $+0$\nu \chi ^{0}$
& 602.7 & 4.60 & 22.0 & 1.54$\times 10^{18}$ & 9.6$\times 10^{20}$ & -- \\ 
&  & 1488.9 & 2.31 & 2.8 & 6.04$\times 10^{18}$ & -- & -- \\ 
$\beta ^{-}\beta ^{-}$; $0_{3}^{+}$ 2153.3 & 0$\nu $+2$\nu $+0$\nu \chi ^{0}$
& 602.7 & 3.19 & 22.0 & 1.07$\times 10^{18}$ & 9.5$\times 10^{20}$ & -- \\ 
&  & 722.8 & 2.04 & 28.0 & 0.54$\times 10^{18}$ & -- & -- \\ 
&  & 827.8 & 2.26 & 26.4 & 0.63$\times 10^{18}$ & -- & -- \\ 
&  & 1550.4 & 0.56 & 5.8 & 0.71$\times 10^{18}$ & -- & -- \\ 
$\beta ^{-}\beta ^{-}$; $2_{5}^{+}$ 2182.4 & 0$\nu $+2$\nu $+0$\nu \chi ^{0}$
& 602.7 & 4.06 & 22.0 & 1.36$\times 10^{18}$ & -- & -- \\ 
&  & 1579.7 & 1.86 & 5.3 & 2.57$\times 10^{18}$ & -- & -- \\ \hline
\end{tabular}
}
\end{center}
\label{tab:thalf-limit}

\end{table*}

\section{Double beta decay study with enhanced sensitivity}

The sensitivity of a future experiment can be enhanced by either increasing the
mass of the sample and the efficiency of the detector or by reducing the background.  
The efficiency of the detector depends on the position and the shape of the sample.  The $\gamma$ rays
can undergo either self-attenuation to loose in the sample itself or  
absorption in other interposed  media. The intensity of the gamma ray decreases to half of its
initial value passing through a thickness of a half-value layer (HVL). The HVL
depends on the atomic number of the element and also on the energy of the $\gamma$
ray. HVL for tin material at
different $\gamma$-ray energies were calculated  from the  mass
attenuation coefficients ($\mu/\rho$) values at different energies taken from NIST
data \cite{nist}.

\begin{figure}[h]
\centering
\includegraphics[width=0.7\columnwidth]{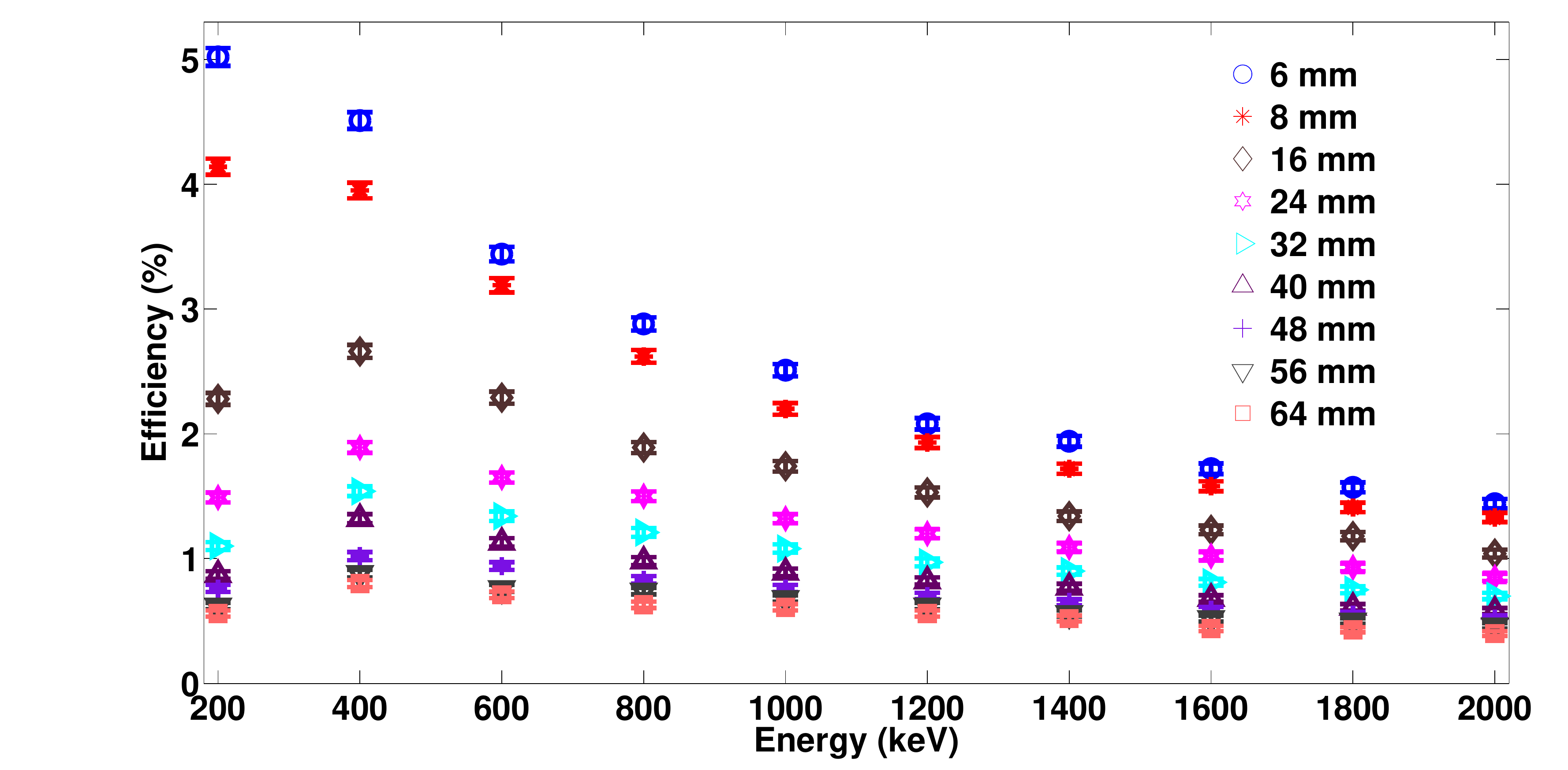}
\caption{Detection efficiency of the HPGe detector at different energies in the region-of-interest 
for different thicknesses of the sample are shown. Legends are the thicknesses of the sample.}
\label{fig:hvl-tin}
\end{figure}
\begin{figure}[h]
\centering
\begin{minipage}[t]{0.45\textwidth}
\centering
\includegraphics[width=\textwidth]{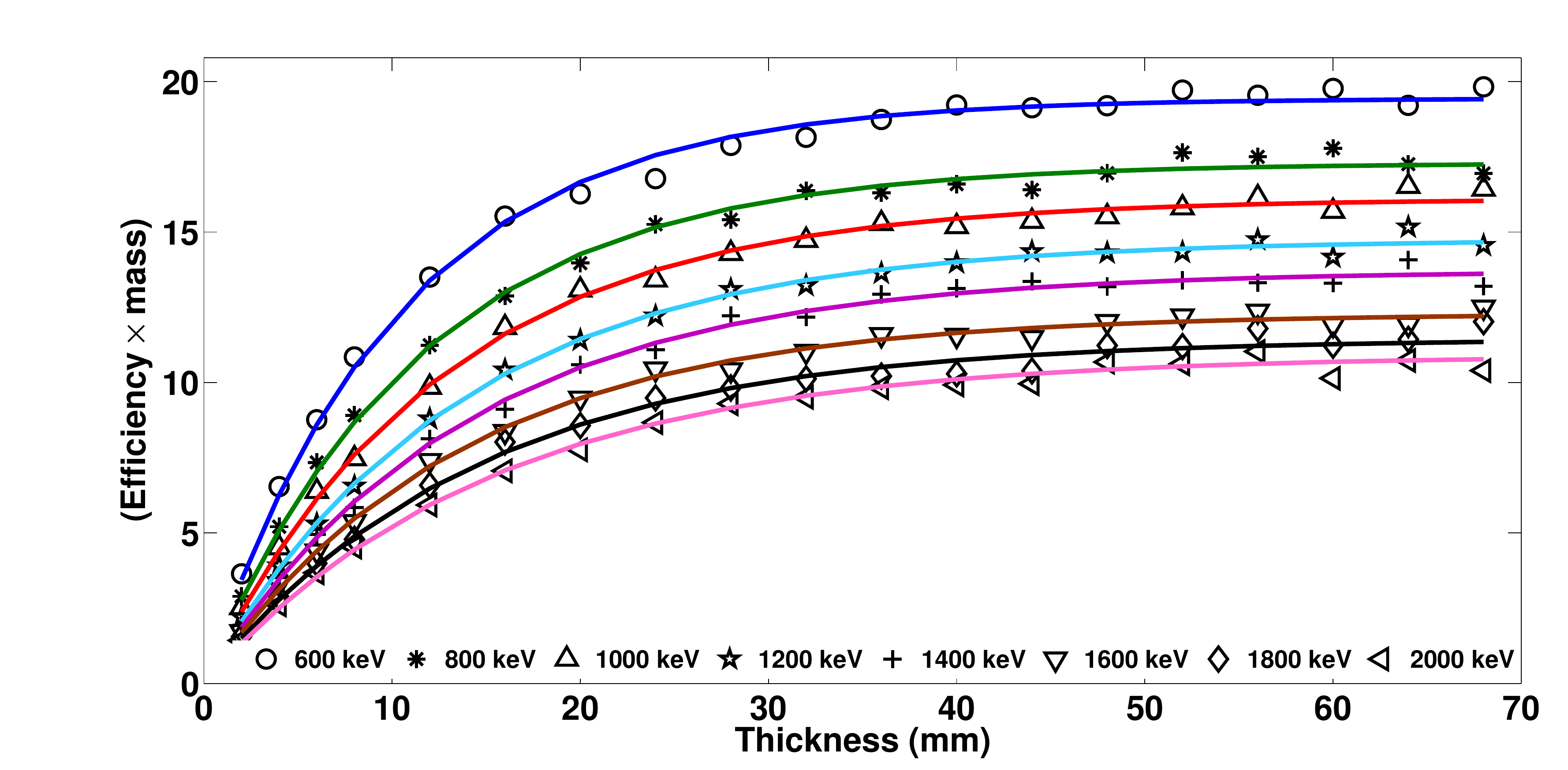}
\caption{Comparison of efficiency times mass of the sample for different thicknesses of the sample.} 
\label{fig:efficiency-mass}
\end{minipage}
\hspace{0.2in}
\begin{minipage}[t]{0.45\textwidth}
\centering
\includegraphics[width=\textwidth]{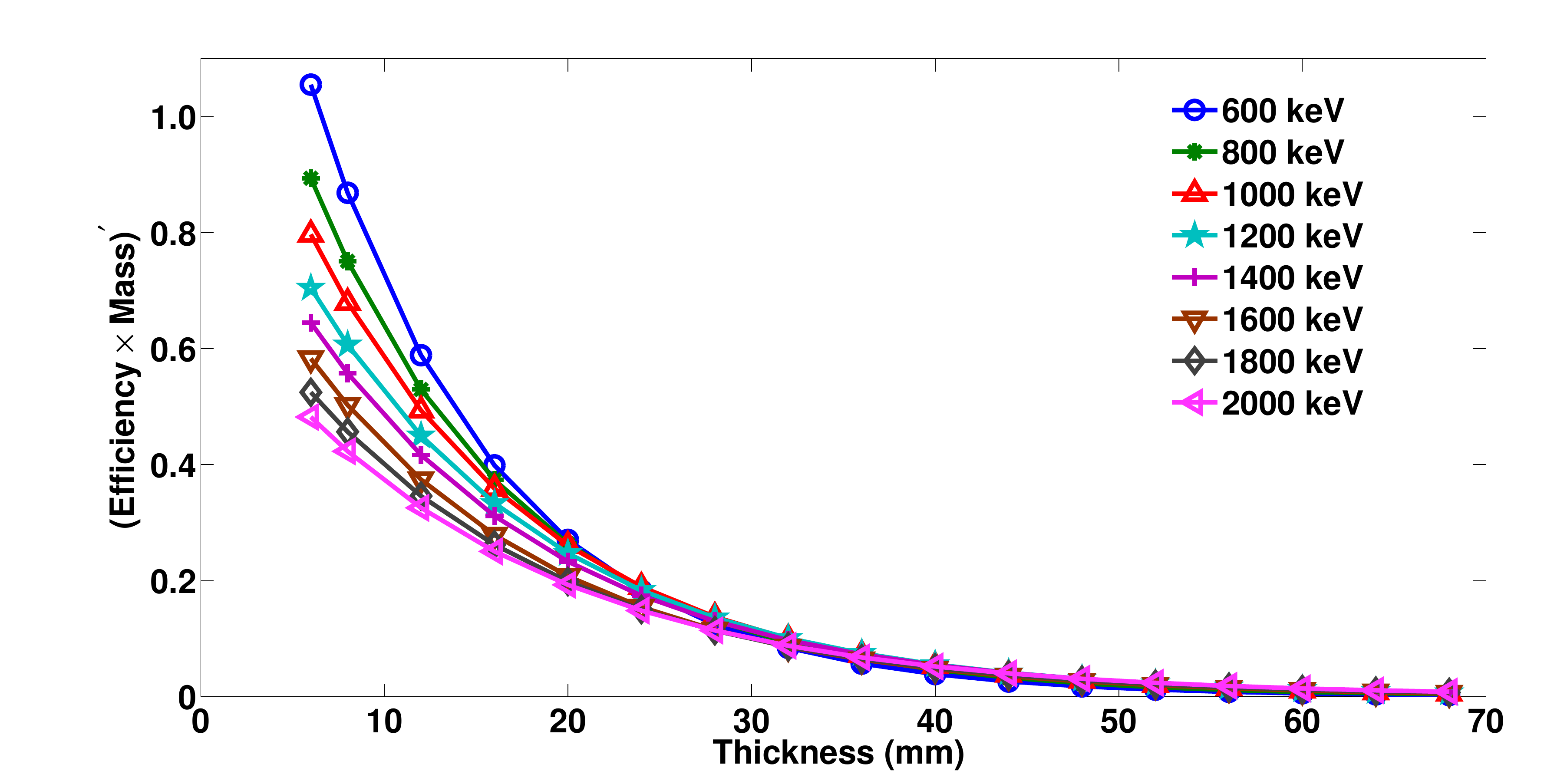}
\caption{Correlation between sample thickness and first derivative of efficiency times mass of the
 sample.}
\label{fig:derivative}
\end{minipage}
\end{figure}

Efficiencies of the detector were simulated for various sample thicknesses and different sample-detector 
distances. At a fixed $\gamma$ ray energy, though the efficiency initially increases for lower energy upto 400 keV and then decreases
with an increasing thickness of the sample as shown in Fig.~\ref{fig:hvl-tin}, the number of double beta nuclei increases 
with increasing the mass of the sample. Hence, the product of the efficiency and the number of 
nuclei together increases with increase in the mass of the sample as shown in Fig.~\ref{fig:efficiency-mass}.

For a cylindrical sample of fixed radius, the growth of the FEP count tends to saturate at a 
certain sample thickness as shown in Fig.~\ref{fig:efficiency-mass}. With further increase 
in sample thickness, the FEP count alters very less but the Compton continuum due to high energy photons
increases. As a result of the increased Compton background, the important low-intensity gamma-lines emitted due 
to double-beta decay processes may not be detected experimentally. Therefore, it is very important to optimized the sample thickness.
The optimization of sample thickness for cylindrical geometry that maximizes the detection efficiency are studied by 
Barrera et al. \cite{barr}, Shweikani et al. \cite{kani} and Li et al. \cite{gang} for HPGe detectors. 
For a  cylindrical sample volume with fixed radius, optimal thickness has been 
determined by calculating the minimum value of the thickness that makes the first derivative of the  sample mass 
multiplied by the detection efficiency nearly zero \cite{kani}. The correlation between
sample thickness and efficiency times mass of the sample can be expressed by the Box Lucas function having the form  \cite{kani} :
\begin{equation}
\eta\cdot M = a(1- e^{-b\cdot h})
\end{equation}  
where $a$, $b$ are two constants and $h$ is sample thickness. The variation of ($\eta\cdot M$) with sample thickness is
shown in Fig.~\ref{fig:efficiency-mass}. In order to estimate the optimum sample thickness at a given energy, the correlation between
the first derivative of efficiency times mass of the sample, ($\eta\cdot M$)$^{\prime}$ and sample thickness was studied.
From Fig.~\ref{fig:derivative}, we can see that the value of ($\eta\cdot M$)$^{\prime}$ tends nearly to zero at sample thickness 
36 mm and higher for  gamma energies 600-2000 keV. We consider $h=36$ mm as optimal thickness for the future large-scale experiment.

\begin{figure}[h]
\includegraphics[width=1.0\columnwidth]{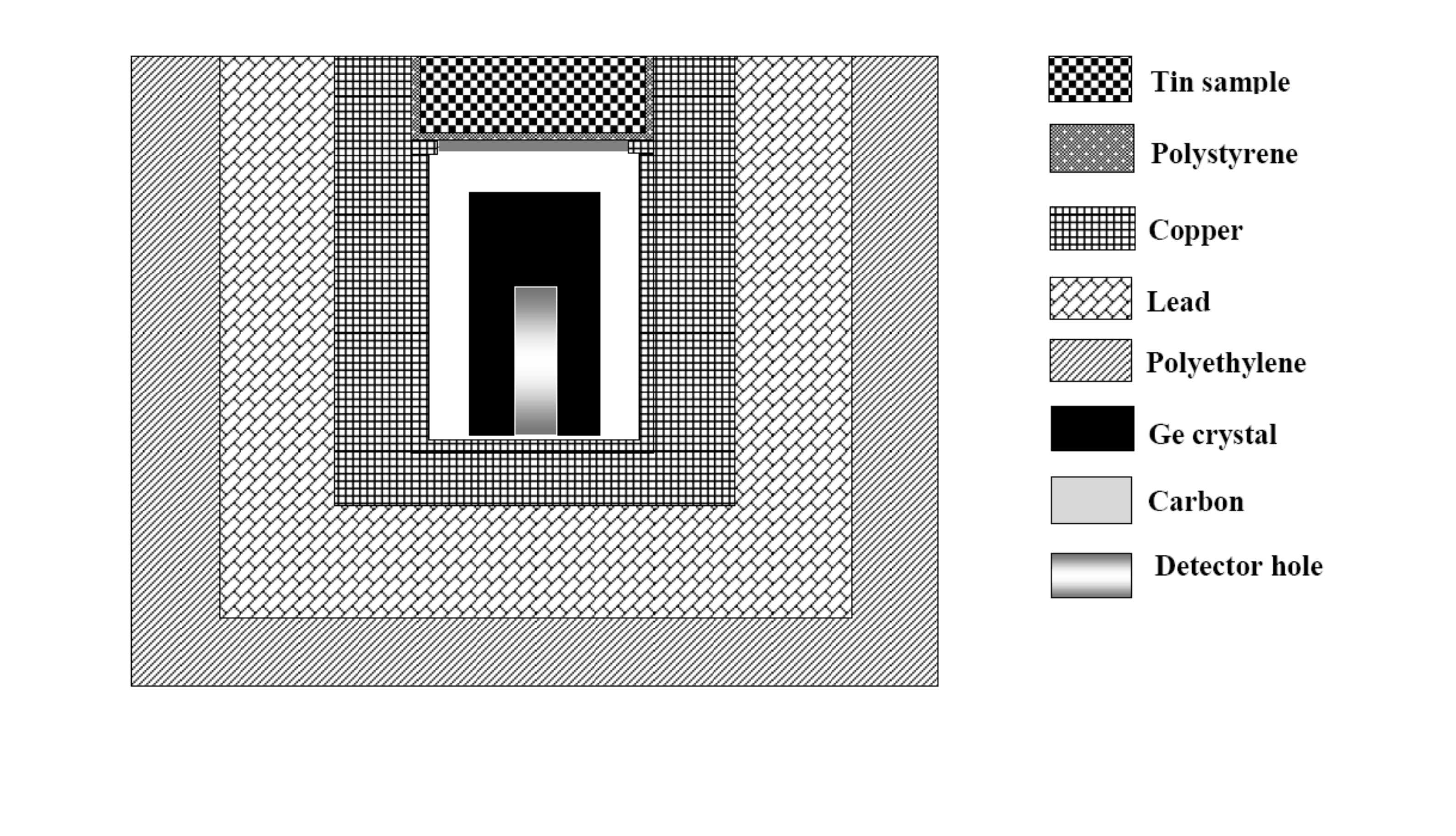}
\caption{Schematic diagram of the experimental set-up of the proposed future experiment. Figure is not in scale.}
\label{fig:hpge2}
\end{figure}
Considering the geometry of the present HPGe detector, a 
cylindrical shaped ($\oslash$86.0 mm $\times$ 36.0 mm)  tin sample of mass  $\sim$1530 g (density of 
tin is taken as 7.31 g/cm$^3$) can be 
placed on the sample holder, as shown in Fig.~\ref{fig:hpge2}.
 A tin sample, enriched in  $^{112}$Sn up-to 90$\%$,
will contain 7.40$\times 10^{24}$ nuclei of this isotope. Enriching in 
$^{124}$Sn instead by the same factor, the sample will contain 6.68$\times 10^{24}$ nuclei of it. 
Considering the efficiencies  2.44 $\%$ (2$\nu$ EC-EC) and 2.36 $\%$ (0$\nu$ EC-EC)
for decay of $^{112}$Sn to the 2$_{1}^{+}$ (617.5 keV) state of $^{112}$Cd, the following sensitivity limits 
(at 90\% C. L.) could be reached after two years of measurements:
$$T_{1/2}^{2\nu EC-EC}(\text{$^{112}$Sn, g.s.} \to \text{617.5 keV} ) \ge 3.0 \times 10^{22} \text{yr,}$$
$$T_{1/2}^{0\nu EC-EC}(\text{$^{112}$Sn, g.s.} \to \text{617.5 keV} ) \ge 3.0 \times 10^{22} \text{yr.}$$
This result would be better than the
previously measured best half-life limit\cite{bash2}. For the double-beta decay transition of
$^{124}$Sn to the excited  2$_{1}^{+}$ (602.7 keV) state of $^{124}$Te, with 90$\%$
enriched $^{124}$Sn and an efficiency of 2.50 $\%$, the  sensitivity would be the same as for $^{112}$Sn, up to $10^{22}$
yr (at 90\% C. L.) for the half-life. 

\section{Conclusion}

In our preliminary study for the search of excited state
transitions concerning possible double beta decays of tin isotopes with a HPGe detector, we measured 
half-life limits of $10^{17} - 10^{18}$ yr 
for $\beta^{+}$EC and EC-EC processes in $^{112}$Sn and
$10^{18}$ yr for $\beta^{-}\beta^{-}$ transition in $^{124}$Sn.
For the  $0\nu$EC-EC decay of $^{112}$Sn to the ground state of its daughter nuclide we obtained  
 a half-life limit of 1.27$\times10^{18}$ yr.  We showed that an 
experiment with larger mass, and material  enriched to high levels of either $^{112}$Sn or $^{124}$Sn, 
could reach with the same HPGe detector half-life limits for both tin isotopes on the order of $10^{22}$ yr for
the decays to the excited levels.
 In India, an effort has been started to build up a bolometric Sn detector
for experimental study of $0\nu\beta\beta$ decay in $^{124}$Sn \cite{vandana}. The experimental
setup will be housed at upcoming India-based Neutrino Observatory site.

\section{Acknowledgement}

The authors  would like to thank Prof. V. I. Tretyak,  Prof. V. Nanal  and Prof. R. G. Pillay for 
fruitful discussions 
and Gran  Sasso laboratory  staff  for  their  technical
assistance  in running  the experiment.  Author Soumik  Das  would  like  to acknowledge 
the  financial assistance from University  Grant Commission, India 
(F  No 10-2(5)/2006(ii)-EUII)  and P.  K. Raina, S.  K. Ghorui  and P.  K. Rath 
acknowledge the financial support of the Council for Scientific and Industrial 
Research, India (CSIR Project No 03(1216)/12/EMR II) and Department of Science and Technology (DST), 
India (SB/S2/HEP-007/2013) for financial assistance during this work. 
Financial supports from DST, India (sanction order no. INT/Italy/p-7/2012(ER)) and MAE, 
Italy (Project Id : IN12MO11) are also acknowledged. We acknowledge the use of the computing facility
from DST-FIST (Phase-II) Project installed in the Department of Physics, IIT Kharagpur, India.

\end{document}